\documentclass[aps,twocolumn,showpacs,preprintnumbers,amsmath]{revtex4}

\usepackage{amssymb}
\usepackage{graphicx}
\usepackage{dcolumn}
\usepackage{bm}

\begin{document}

\title{Lightning Stars: Anomalous Photoproduction in Neutron Stars via Parametric Resonance Mechanism}

\author{S. D. Campos} \email{sergiodc@ifi.unicamp.br}
\affiliation{Instituto de F\'isica Gleb Wataghin, Universidade
Estadual de Campinas, 13083-970, Campinas, SP, Brazil.}

\author{A. Maia Jr.} \email{maia@ime.unicamp.br}
\affiliation{Departamento de Matem\'{a}tica Aplicada, Universidade
Estatual de Campinas, 13083-970, Campinas, SP, Brazil.}

\date{\today}

\begin{abstract}
In this work we propose a new mechanism for photoproduction inside
a neutron star based on Parametric Resonance phenomenon as firstly
applied to Inflationary Cosmology. Our assumptions are based on
the pion condensation model by Harrington and Shepard. We show
that a huge number of photons are created which, on turns, reheats
the matter in the star's core. Thus, we argue that Parametric
Resonance can be effective during a brief period out of an neutron
star lifetime leading to an anomalous uprising variation of its
brightness departing from the black body radiation at regularly
spaced frequencies. In adition, a time periodic signal is obtained in 
moderate (not exponential) regimes. We argue also that our PR mechanism offers a 
simple and feasible explanation for some  recent observations of 
giant flares from neutron stars. 
\end{abstract}

\pacs{97.60.Jd,13.60.L}

\maketitle

\section{Introduction}

It is well known that in Inflationary Cosmology the early Universe
turns out be depleted of its matter content at the end of
inflation era. So a mechanism is necessary in order to fill up the
Universe with baryonic matter (baryogenesis), as well dark matter.
In current models, matter must be generated at a very high
temperature, after the inflation era, in order to start
baryogenesis. A mechanism, firstly proposed by Traschen and
Brandenberger \cite{tb} and independently by Dolgov and Kirilova
\cite{dk} based on Parametric Resonance (PR) phenomenon
\cite{lan}, succeed to provide such requirements as a possible way
to generate matter in pos-inflationary Universe with an
exponential particle production. Few years later, Kofman
\textit{et al} \cite{kls} showed that this mechanism is effective
mainly in the so-called preheating phase of the inflation era.

The simplest model in Field Theory consists of two boson fields,
namely, a self-interacting source field $\phi$ and a resonant (or
matter) field $\chi$. From the mathematical point of view, the
source field $\phi$ can be obtained as solution of an ordinary
second order nonlinear differential equation. Periodic solutions
are founded when the field $\phi$ is in the oscillatory regime
around the minimum of its potential (for example, in
$\lambda\phi^4$ model). In the simplest (spatial homogeneous)
case, one looks for periodic solutions which in turns enter in the
ordinary linear (time) differential equation of the matter field
$\chi$ as a periodic potential. These linear equations are
particular cases of the so called Hill's Equation \cite{win,
arscott}. The important point for these models is that Hill's
Equation presents, in its parameter space, instability regions,
also named resonant bands. In a resonance band the amplitude of a
$\chi$ solution can grow exponentially and, since the particle
number is proportional to this amplitude one obtains a huge number
of $\chi$ quanta \cite{kls}. If $\chi$ is a boson field, the
particle number occupation with a specific momentum is
exponentially large. On the other hand, Dolgov and Kirilova
\cite{dk} showed that the PR mechanism is not very effective
(particle production is not exponential) for fermions field due to
Pauli blocking and the solutions presents only a polynomial
growth.

Few years ago, Garc\'{\i}a-Bellido and Kusenko \cite{gbk} proposed
that the same PR mechanism could be used to explain the huge
energy coming from "point-like" sources in the sky, i.e. the
so-called cosmic ray bursts. In order to obtain the needed
potential, bounded from below, the above authors took the
collision of two neutron stars (NS) which, merging in each other
produce a very compact object whose proton matter inside turns out
to be superconductor and can be modelled by a Ginzburg-Landau
potential \cite{landau}. In addition, they took electromagnetic
field as a resonant field interacting with the superconductor
matter. So, when the system is in the PR regime a huge burst of
energy is transferred from superconductor matter to the
electromagnetic field in form of high energy photons and these, in
principle, could be detected on the Earth as cosmic ray bursts. As
far as we known this was the first application of the PR mechanism
in order to explain the bursts coming from compact objects in the
sky.

Nevertheless we wonder whether PR can be applied, not to the
spectacular collision of two NS, but to a single star itself in a
more moderate energy scale, of course. This was motivated, after
Walecka's pioneering work \cite{wal} on quantum field theory
applied to stellar structure. So, differently from the above cited
researchers, we apply the PR mechanism to a single NS, whose part
of its matter content in the core has supranuclear densities. We
are mainly interested in the regime in which matter in the star
medium is populated with negative pion described by a field in a
condensate phase \cite{glen} which oscillates around the minimum
of its potential as required by the PR mechanism. This oscillation
is a natural consequence of the energy excess coming from the
formation of the pion condensate, that is, the system to turns in
a condensate phase must reduces its energy and therefore is
plausible to suppose that this excess is then transferred to the
condensate field causing its motion. As in Garc\'{\i}a-Belido and
Kusenkos's  model, the role of resonant field is played by the
electromagnetic field interacting with this oscillatory field.

We shall show below that the proposed model leads to a very
efficient, although not thermal, photoproduction. We expect that
in a second phase (quite the same as in Inflationary Cosmology)
the production will thermalize with the surrounding matter and
consequently changing the equation of state (EoS) of matter inside
the star. Of course, a careful study of energy transport is needed
in order to evaluate the impact of this kind of photoproduction to
the star evolution. In this work we just show that Parametric
Resonance is a plausible mechanism for the pion-photon model
leading to an anomalous and fast energy production in the interior
of compact stars.  The new, nevertheless important, questions
above mentioned deserve, of course, a careful further study.

\section{Pion Condensation in Neutron Stars}

Neutron stars have provided us with information on the properties
of high density nuclear matter. Accumulated data have given useful
information about reliable EoS and thermal properties of the NS
nuclear matter which seem to suggest the appearing of some phase
transitions there, e.g, pion condensation. The EoS then can be
used to determine NS parameters, most notably the mass-radius
relationship and the maximum allowable mass \cite{shapiro}.

An interesting phase transition that can take place inside a NS is
the pion condensation. This possible condensation is due, to a
large extent, to the high density of the neutron star medium. The
possibility of pion condensation in nuclear or neutron matter was
initially proposed by Migdal \cite{mig}, Sawyer \cite{saw} and
Scalapino \cite{sca}. Whether pion condensate exist or not is of
great interest since it can significantly enhance the cooling
rates of NS \cite{kog,max}. A very fast cooling for NS may be
explained by a possible pion condensation \cite{nomoto}.
PR mechanism start to work just after the condensate is formed as we
shall show in the following sections.

Condensed phase of negative pion ($\pi^-$) can take place if we
neglect the strong pion correlations with ordinary matter in
modifying the pion self energy. It is possible to show (see ref.
\cite{mig1}) that the favorable condition to this assumption is
satisfied for a neutron on the top of the Fermi sea to turn a
proton and $\pi^-$ when

\begin{eqnarray}
\nonumber {\mu_n}-{\mu_p}={\mu_e}>{m_{\pi}},
\end{eqnarray}

\noindent where $m_{\pi}{\simeq}140$ $\mathrm{MeV}$ is the $\pi^-$
rest mass. When this occurs, we say that the negative pion
macroscopically occupy a single mode in its ground state, that is,
they are in a condensate state.

The $\pi^0$ condensation can take place also, since its effective
mass in the medium is zero \cite{mig1}. Thus, the decay reaction
occurs

\begin{eqnarray}
\nonumber n{\rightarrow}n+\pi^0.
\end{eqnarray}

Possible condensation of positive pion ($\pi^+$) will not taken
into account here since Sawyer and Yao \cite{sy} had showed that
in NS medium the number of $\pi^+$ particles is much smaller than
the $\pi^-$ particles. Therefore, even when $\pi^+$ condensation
occurs it could be neglected when compared with the $\pi^-$
condensation. Neutral pion does not interact with the
electromagnetic field of the NS and the $\pi^+$ condensate will
does not affect the electromagnetic field in significant way,
considering the above argument. So, for the photoproduction, in
our case, we can take into account only the interaction between
the negative pion condensate and the electromagnetic field of the
NS. Henceforth, we will indistinctly use the wording pion
condensate as $\pi^-$ condensate.

\section{General Formalism in the Sigma Model}

In order to get a $\pi^-$ condensate we will use the Lagrangian
formalism for the linear $\sigma$ model. Although the linear
$\sigma$ model is a toy model it is enough, for our purposes,
to do a qualitative study of the PR phenomenon.

We start by considering the
Lagrangian density for the electromagnetic field $A^{\mu}$, the
isoscalar $\sigma$, the isotriplet $\vec{\pi}$, and the nucleon
isodoublet $\Psi$ \cite{hs1,hs2}:

\begin{equation}
\label{1}\mathcal{L}=\mathcal{L_A}+\mathcal{L_\sigma}+\mathcal{L_{\pi^-\pi^+}}+
\mathcal{L}_{\pi^0}+\mathcal{L}_{\Psi}-\mathcal{L}_{int},
\end{equation}

\noindent where

\begin{eqnarray}
\nonumber \mathcal{L_A}=
-\frac{1}{4}F_{\mu\nu}F^{\mu\nu}+e^2(A^{\mu})^2,
\end{eqnarray}

\noindent is the electromagnetic field contribution,

\begin{eqnarray}
\nonumber
\mathcal{L_\sigma}=\frac{1}{2}(\partial_{\mu}\sigma)^2+\frac{1}{2}{m_0}^2{\sigma}^2-\frac{\lambda^2}{4}{\sigma}^4+{c_1}\sigma,
\end{eqnarray}

\noindent is the $\sigma$ contribution. Note that $c_1{\sigma}$ term
is responsible for the spontaneous symmetry breaking in this model.

\begin{eqnarray}
\nonumber
\mathcal{L_{\pi^-\pi^+}}=\partial_{\mu}\pi^-\partial^{\mu}\pi^++\\
\nonumber+\frac{1}{2}{m_0}^2(2\pi^-\pi^+)-\frac{\lambda^2}{4}(2\pi^-\pi^+)^2,
\end{eqnarray}

\noindent is the charged pion contribution,

\begin{eqnarray}
\nonumber
\mathcal{L}_{\pi^0}=\frac{1}{2}(\partial_{\mu}\pi^0)^2+\frac{1}{2}{m_0}^2{\pi^0}^2-\frac{{\lambda}^2}{4}{\pi^0}^2,
\end{eqnarray}

\noindent represents the neutral pion contribution,

\begin{eqnarray}
\nonumber
\mathcal{L}_{\Psi}=\bar{\Psi}[i\gamma_{\mu}\partial^{\mu}]{\Psi},
\end{eqnarray}
\noindent denote the nucleon contribution and finally, the
interaction Lagrangian density reads

\begin{widetext}
\begin{eqnarray}
\nonumber \mathcal{L}_{int}=-ie(\pi^+\partial_{\mu}\pi^--\pi^-\partial_{\mu}\pi^+)A^\mu-\bar{\Psi}[e\gamma_{\mu}A^{\mu}\frac{1}{2}(1+\tau_3)+-g(\sigma+i\vec{\tau}\cdot{\vec{\pi}}\gamma_5)]\Psi+\\
\nonumber+\frac{{\lambda}^2}{4}[{\sigma^2}{\pi^0}^2+{\sigma}^2(2\pi^-\pi^+)+{\pi^0}^2(2\pi^-\pi^+)],
\end{eqnarray}
\end{widetext}

\noindent where $\tau_j=1,2,3$ are the Pauli matrices and

\begin{displaymath}
\gamma_5=-\left(\begin{array}{c}
0\hspace{0.3cm}1\\
1\hspace{0.3cm} 0\\
\end{array}\right).
\end{displaymath}

The nucleon isodoublet $\Psi$ contains four-component proton
and neutron spinor fields and can be written as

\begin{displaymath}
\Psi=\left(\begin{array}{c}
\Psi_p\\
\Psi_n\\
\end{array}\right),
\end{displaymath}

\noindent and $\pi$ denotes the pion field operator
that annihilates a $\pi^-$ (or creates a $\pi^+)$, which we can write as

\begin{eqnarray}
\label{2} \pi^{\pm}=\frac{\pi_1 \pm i \pi_2}{\sqrt{2}},
\end{eqnarray}

\noindent where $\pi_1$ and $\pi_2$ are real components of the pion field and
$\vec{\pi}=(\pi^{0},\pi^{+},\pi^{-})$.

The constants in the Lagrangian density can be related (in tree
approximation) to the physical quantities $f_{\pi}{\simeq}93$
$MeV$, $m_{\pi}{\simeq}140$ $\mathrm{MeV}$ \cite{camp1,camp2}:

\begin{eqnarray}
\nonumber c_1=f_{\pi}{m_{\pi}}^2,\\
\nonumber 2{\lambda}^2{f_{\pi}}^2={m_{\sigma}}^2-{m_{\pi}}^2,\\
\nonumber 2{m_0}^2={m_{\sigma}}^2-3{m_{\pi}}^2.
\end{eqnarray}

Note that if we add a symmetry breaking term in the $\pi$
Lagrangian density, instead in $\sigma$ Lagrangian, then different
constants must be defined \cite{camp1}. Finally
$e=\frac{4\pi}{137}$ is the electromagnetic coupling constant and
$g=f_{\pi}m$, where $m$ is the nucleon mass (typically
$m{\simeq}940$ $\mathrm{MeV}$). Usually the mass of $\sigma$ meson
is taken as bigger than 1 $\mathrm{GeV}$ \cite{sch,gell}. Without
loss of generality, we take $m_{\sigma}=1020$ $\mathrm{MeV}$, and
then we get $m_0{\simeq}700$ $\mathrm{MeV}$, $\lambda{\simeq}7.7$
and $g{\simeq}87420$ $\mathrm{MeV}^2$. \textit{A priori} we do not
have information about the $\sigma$  meson and in the NS medium we
have a lot of candidates to be the ``$\sigma$ meson'' \cite{glen}.
Since this is a qualitative study we choose this value for
$m_{\sigma}$ due the mass of $\phi$ meson, whose mass is around
$1019$ $\mathrm{MeV}$.

Harrington and Shepard \cite{hs1,hs2}, using (\ref{1}), showed
that the pion condensate and the strong electromagnetic field of a
NS may co-exist in a particular region (shell) inside of the
compact star, as well nucleons. They showed also that the pion
condensate can assume a superconducting behavior, even in the
presence of an intense electromagnetic field. According, we assume
in this work that the pion condensate and the electromagnetic
field may co-exist inside of the NS, as showed by the authors of
\cite{hs1,hs2}. In terms of Harrington and Shepard's work the pion
condensation, without a superconducting phase transition, occurs
when the neutron star magnetic field $H$ is weaker than some
critical low value, $H_{c1}$. This sets the basis for our physical
model.

Now, Parametric Resonance phenomenon belongs to a class of
ordinary linear second order differential equation called Hill's
Equation \cite{win}. This class possess two distinct important
cases (but they are not the only ones, of course): the Mathieu's
and Lam\'{e}'s ones. In the context of particle production by the
PR mechanism, Mathieu's Equation was firstly used by Traschen and
Brandenberger \cite{tb} and Dolgov and Kirilova \cite{dk} and
Lam\'{e}'s Equation by Kofman \textit{et al} \cite{kls}. Following
the above authors we find the equation of motion (EoM) for the
pion condensate and its periodic solutions which, in turn, enter
as periodic potentials in the EoM for the electromagnetic field.
These are Hill's Equations whose the most useful ones are the
Mathieu and Lam\'{e} types as mentioned above.

The specific EoM for $A^{\mu}$ depends only on the form of
periodic solution for the pion condensate EoM. If its periodic
solution is a simple trigonometric function then the $A^{\mu}$'s
EoM is classified as Mathieu's type; if its solution is an
elliptic function, the EoM is classified as Lam\'{e}'s type. For
both types, solutions for $A^{\mu}$'s EoM are guaranteed by a
Floquet's theorem \cite{win} and the behavior (growth) of a
particular solution is determined, basically, by its Floquet
exponent. If the Floquet exponent is a real number, the solutions
are unstable, i.e., they  present resonant bands in their
parameter space in which the solutions growth (or decay)
exponentially. In the case of imaginary number, the solutions are
stable (bounded).

So, in order to find the EoM for the electromagnetic field we must
find firstly the EoM to the $\pi^-$ condensate, applying the
Euler-Lagrange equations in (\ref{1}). We get,

\begin{widetext}
\begin{eqnarray}
\label{3}
\Box{\pi^-}+\left[-{m_0}^2+\frac{\lambda^2}{2}\sigma^2+\frac{\lambda^2}{2}{\pi^0}^2\right]{\pi^-}+\lambda^2{\pi^+}(\pi^-)^2+ig\sqrt{2}\bar{\Psi}[{\tau_-}\gamma^5]\Psi=0,
\end{eqnarray}
\end{widetext}

\noindent where $\tau_-=(\tau_1-i\tau_2)/{\sqrt{2}}$ and the back
reaction of $A^{\mu}$ field over $\pi^-$ condensate was neglected.
As mentioned above, our purpose is to study the effects of $\pi^-$
condensate over the $A^{\mu}$ field and then investigate the PR
phenomenon inside of an isolated NS. Back reaction of $A^{\mu}$
field can affect the evolution of $\pi^-$ condensate, but we
expect, in a first order approximation, that this contribution may
be neglected as in the preheating phase of the Inflationary
Cosmology \cite{tb, kls} .

The $\pi^0$ field does not interact with the $A^{\mu}$ field, so
we can take it in its ground state

\begin{eqnarray}
\nonumber \langle{\pi^0}\rangle_{vac}=0
\end{eqnarray}

\noindent and this implies a non condensation phase to $\pi^0$
meson \cite{mig1}. The $\pi^-$ condensate is not affected by this
assumption since conditions on both condensations are independent,
as seen above. Now taking the approximation
$\langle{\pi^0}^2\rangle{\simeq}\langle{\pi^0}\rangle^2$ we can
neglect the $\pi^0$ contribution in equation (\ref{3}).

Symmetry breaking term presents in $\mathcal{L}_{\sigma}$
(equation (\ref{1})) leads to a $\sigma$ condensation prevent us
to consider $\langle{\sigma}\rangle_{vac}=0$ \cite{camp1,camp2}.
In order to estimate the $\sigma$ contribution we take
$\langle{\sigma}\rangle=b \langle{\sigma}\rangle_{vac}$, where $b$
is a positive real number. Following \cite{camp1,camp2,hs1,hs2} we
may write

\begin{eqnarray}
\nonumber \langle{\sigma}\rangle_{vac}{\simeq}f_{\pi}\cos\theta
\end{eqnarray}

\noindent where $\theta{\neq}0$ is the chiral symmetry angle that
gives information about the $\sigma$ and $\pi$ condensation. So,

\begin{eqnarray}
\nonumber \langle{\sigma}\rangle=b\langle{\sigma}\rangle_{vac}{\simeq}af_{\pi},
\end{eqnarray}

\noindent where $a=b\cos\theta>0$ is a real parameter.

Then, equation (\ref{3}) reads, after some algebra with the nucleon components,

\begin{widetext}
\begin{eqnarray}
\label{4}\Box{\pi^-}+\left[\frac{\lambda^2{f_\pi}^2{a^2}}{2}-{m_0}^2\right]{\pi^-}+\lambda^2{\pi^+}(\pi^-)^2-ig\sqrt{2}{|\Psi_n|}^2=0.
\end{eqnarray}
\end{widetext}

If we take values previously obtained for $m_0$, $\lambda$ and
$f_{\pi}$, we obtains two distinct cases for equation (\ref{4}):

\begin{center}
$a^2<\frac{2{m_0}^2}{\lambda^2{f_{\pi}}^2}{\simeq}1.92$,

$a^2>\frac{2{m_0}^2}{\lambda^2{f_{\pi}}^2}{\simeq}1.92$.
\end{center}

For the sake of simplicity we can take into account these two
inequalities defining

\begin{eqnarray}
\nonumber
\frac{{\lambda^2}{f_{\pi}}^2{a^2}}{2}-{m_0}^2=\pm2M^2,
\end{eqnarray}

\noindent where we have case $-2M^2$ corresponds to $a^2<1.92$ and
case $+2M^2$ to $a^2>1.92$. So we can write equation (\ref{4}) as

\begin{eqnarray}
\label{5}\Box{\pi^-}\pm2M^2{\pi^-}+\lambda^2{\pi^+}(\pi^-)^2-ig\sqrt{2}{|\Psi_n|}^2=0.
\end{eqnarray}

Also, as a first approximation, we can consider here a region
inside the NS in which $\pi^-$ condensate has a homogeneous and
isotropic distribution. So, we can neglect spatial contributions
and then equation (\ref{5}) may be rewritten as

\begin{eqnarray}
\label{6}\frac{d^2}{dt^2}{\pi^-}\pm2M^2{\pi^-}+\lambda^2{\pi^+}(\pi^-)^2-ig\sqrt{2}{|\Psi_n|}^2=0.
\end{eqnarray}

We will analyze the above equation (with $\pm$ signs) in two
cases: Homogeneous (${|\Psi_n|}^2=0$) and Inhomogeneous
(${|\Psi_n|}^2{\neq}0$) equation. We will see that for both signs
we get periodic solutions which lead to PR for the electromagnetic
field .

\subsection{Homogeneous Case ( ${|\Psi_n|}^2=0$)}

Neglecting nucleon contributions we can write (\ref{6}) as

\begin{eqnarray}
\label{7}\frac{d^2}{dt^2}{\pi^-}\pm2M^2{\pi^-}+\lambda^2{\pi^+}(\pi^-)^2=0.
\end{eqnarray}

This ordinary nonlinear second order differential equation is not
in any way a simple one. It can be solved in terms of
Weierstrass's Elliptic Functions \cite{ince} which  are periodic. We can use
relation (\ref{2}) and write (\ref{7}) as a real ordinary
nonlinear second order differential coupled system:

\begin{equation}
\left\{\begin{array}{r@{\quad\quad}}
\label{8}\frac{d^2}{dt^2}\pi_1\pm{2M^2}\pi_1+\frac{\lambda^2}{2}{\pi_1}^3+\frac{\lambda^2}{2}{\pi_2}^2{\pi_1}=0\\
\frac{d^2}{dt^2}\pi_2\pm{2M^2}\pi_2+\frac{\lambda^2}{2}{\pi_2}^3+\frac{\lambda^2}{2}{\pi_1}^2{\pi_2}=0,
\end{array}\right.
\end{equation}

\noindent where the same sign must be taken simultaneously in both equations  for the
factor $2M^2$.

It is easy to see that system described by (\ref{8}) is symmetric
under $\pi_1{\longleftrightarrow}\pi_2$ exchange. This fact
suggests similarities between amplitudes, wavelength and
frequencies of $\pi^-$ and $\pi^+$ solutions. In the next subsection we will
make a numerical analysis of this system.

\subsubsection{Numerical Solution}

\begin{figure*}
\centering{\includegraphics[scale=0.7]{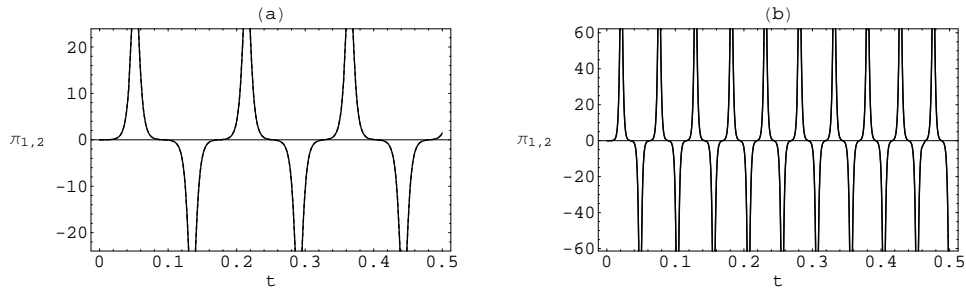}}
\caption{\label{fig:1}For the case $-2M^2$: (a) $M=140$
$\mathrm{MeV}$ and (b) $M=420$ $\mathrm{MeV}$. Note that $\pi_1$
and $\pi_2$ are similar in (a)
  and (b). In both cases, $\lambda=7.7$.}
\end{figure*}

Figure \ref{fig:1} shows the evolution of the system (\ref{8}) for
the case $-2M^2$ with $M=140$ $MeV$ and $M=420$ $MeV$. These
values were obtained when $a^2=1.78$ and $a^2=0.54$, respectively.
Since $a^2<1.92$, these values can be completely arbitrary
(recall, $a>0$ always). We took some values for $M$ as one pion
mass and three pion mass, approximately, just for numerical
analysis. \textit{A priori}, any real positive value for $M$ may
be used.

\begin{figure*}
\centering{\includegraphics[scale=0.7]{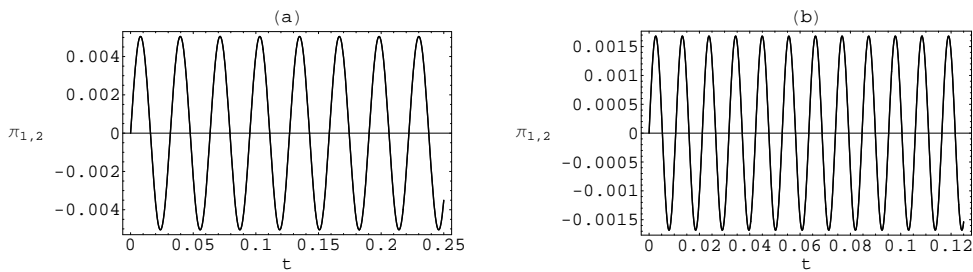}}
\caption{\label{fig:2}For the case $+2M^2$: (a) $M=140$
$\mathrm{MeV}$ and (b) $M=420$ $\mathrm{MeV}$. In both cases,
$\lambda{\simeq}7.7$. Again, they are
 similar in (a) and (b).}
\end{figure*}

Figure \ref{fig:2} shows the temporal evolution of the system
(\ref{8}) for the case $+2M^2$ for $M=140$ $MeV$ and
$M=420$ $MeV$. These values for $M$ were obtained when $a^2=2.1$
and $a^2=3.3$, respectively.

In both Figures, \ref{fig:1}  and \ref{fig:2}, the auxiliary
$\pi_1$ and $\pi_2$ have a periodic behavior.
Therefore, as a direct consequence of this fact, the pion
condensate will have a periodic behavior by (\ref{2}). But it is
still necessary to know what kind of regularity we have: a
trigonometric one or an elliptic one.

\begin{figure*}
\centering{\includegraphics[scale=0.5]{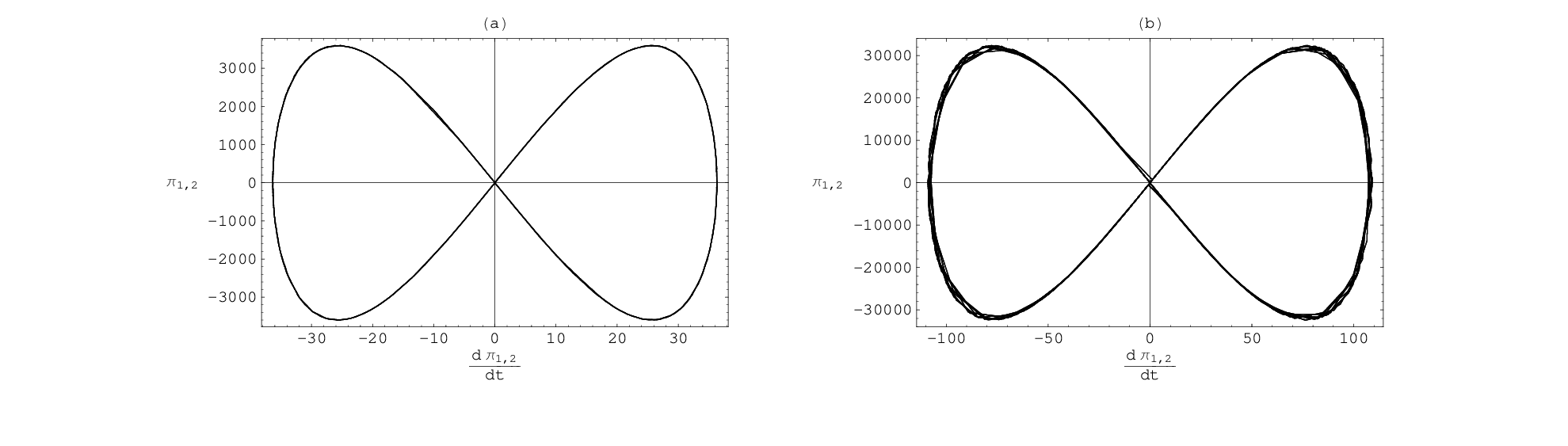}}
\caption{\label{fig:3}Phase space for the case $-2M^2$: (a)
$M=140$ $\mathrm{MeV}$ and (b) $M=420$ $\mathrm{MeV}$. In both
cases, $\lambda=7.7$. The shape of these curve suggests periodic,
bounded solutions as elliptic functions, as can be seen in
\cite{struble}.}
\end{figure*}

\begin{figure*}
\centering{\includegraphics[scale=0.5]{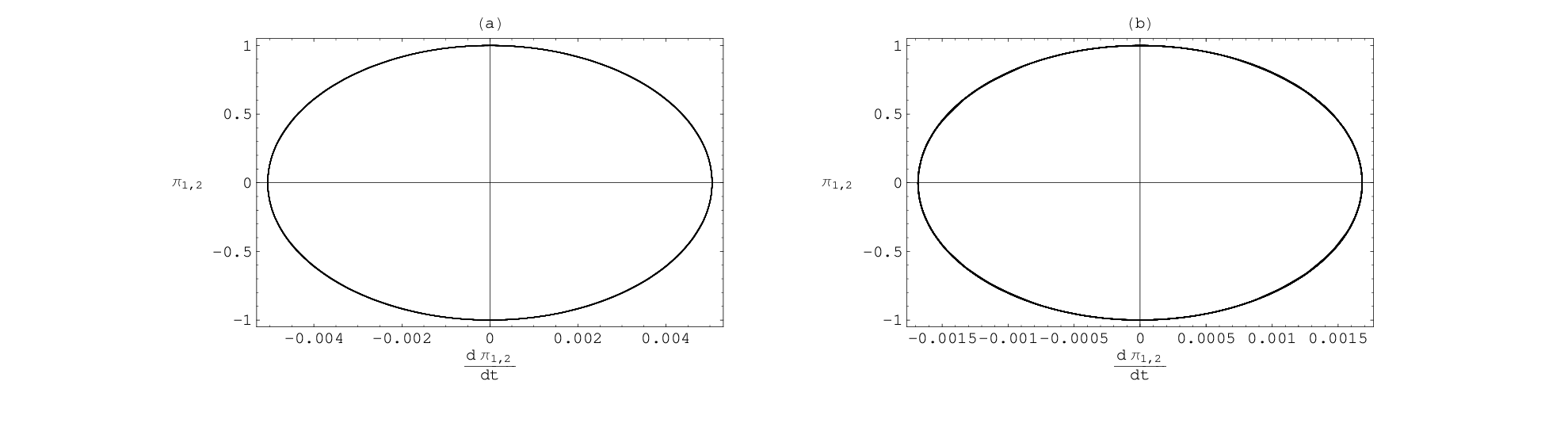}}
\caption{\label{fig:4}Phase space for the case $+2M^2$: (a)
$M=140$ $\mathrm{MeV}$ and (b) $M=420$ $\mathrm{MeV}$. In both
cases, $\lambda=7.7$. Also the elliptic
  behavior is suggested by these graphs.}
\end{figure*}

Figure \ref{fig:3} shows the phase space
$(\dot{\pi}_{1,2},\pi_{1,2})$ for the case $-2M^2$. Figure
\ref{fig:4} shows the phase space for
$(\dot{\pi}_{1,2},\pi_{1,2})$ for the case $+2M^2$. Both pictures
confirm the periodic behavior of solutions for $\pi_1$ and $\pi_2$
and suggest us that periodic solutions shown in Figure \ref{fig:1}
and \ref{fig:2} have an elliptic function behavior, that is, they
could be described by Jacobi's Elliptic Functions. Therefore, as
we will see, Hill's Equation that comes out from (\ref{1}) for the
$A^{\mu}$'s EoM is, in fact, a Lam\'{e}'s Equation whose solutions
are of elliptic type.

It must be observed that (\ref{8}) is an ordinary second order
differential nonlinear coupled system and this may imply a strong
dependence on a given set of initial conditions. Alongside this
paper we adopt, without loss of generality, ${\pi}_1(0)=0$ and
$\dot{\pi}_1(0)=1$ (the same values for $\pi_2$, of course).

\begin{table}
\centering
\begin{tabular}{c|c|c|c}
\hline
Case & Mass $\mathrm{MeV}$ & Wavelength $(l)$ & Frequency $(\nu)$\\
\hline
$(-2M^2)$ &$ M=140$ &${\sim}$ 0.1499 &${\sim}$ 6.6675 \\
$(-2M^2)$ &$ M=420$ &${\sim}$ 0.0541 &${\sim}$ 18.4942 \\
$(+2M^2)$ &$ M=140$ &${\sim}$ 0.0317 &${\sim}$ 31.5111 \\
$(+2M^2)$ &$ M=420$ &${\sim}$ 0.0106 &${\sim}$ 94.5337 \\
\hline
\end{tabular}
\caption{Table with wavelength and frequency for $M=140$
$\mathrm{MeV}$ and $M=420$ $\mathrm{MeV}$ in $-2M^2$ and $+2M^2$
cases.}
\end{table}

It is possible to obtain, numerically, an estimative of wavelength
($l$) and frequency ($\nu$) for the cases $-2M^2$ and $+2M^2$, if
we consider roughly $\nu{\simeq}l^{-1}$. The Table I show some
values of $l$ and $\nu$.

Particle production by the PR mechanism depends strongly on the
amplitude and frequency oscillation of the source field, that is,
on the behavior of the $\pi^-$ field. Determination of these
parameters provide a roughly quantitative estimate of the
particles number produced in the NS medium.

\subsubsection{Exact Solutions}

The symmetry of system (\ref{8}) and its numerical analysis
suggest us the ansatz

\begin{eqnarray}
\nonumber \pi_2=\pi_1,
\end{eqnarray}

\noindent and the system is now reduced to a single ordinary nonlinear
differential equation written as

\begin{eqnarray}
\label{9}\frac{d^2}{dt^2}\pi_1\pm{2M^2}\pi_1+{\lambda^2}{\pi_1}^3=0.
\end{eqnarray}

The above equation is the so-called Duffing's Equation
\cite{struble} and can be solved in terms of Jacobi's Elliptic
Functions. This fact is in complete agreement with the above
numerical analysis. The Duffing's theory allow us to understand
the behavior of exact solutions as follow below.

Equation (\ref{9}) can be transformed in a following system:

\begin{eqnarray}
\left\{\begin{array}{l@{\quad\quad}}
\nonumber\frac{d\pi_1}{dt}=\eta\\
\frac{d\eta}{dt}=-{\lambda^2}{\pi_1}^3{\pm}{2M^2}\pi_1,
\end{array}\right.
\end{eqnarray}

\noindent where $\eta=\eta(t)$ is a real function. The field
direction of this system in the phase plane $(\eta,\pi_1)$ can be
written as

\begin{eqnarray}
\nonumber \frac{d\eta}{dt}\frac{dt}{d\pi_1}=\frac{d\eta}{d\pi_1}=-\frac{\lambda^2{\pi_1}^3{\pm}2M^2{\pi_1}}{\eta}.
\end{eqnarray}

\begin{figure*}
\centering{\includegraphics[scale=0.9]{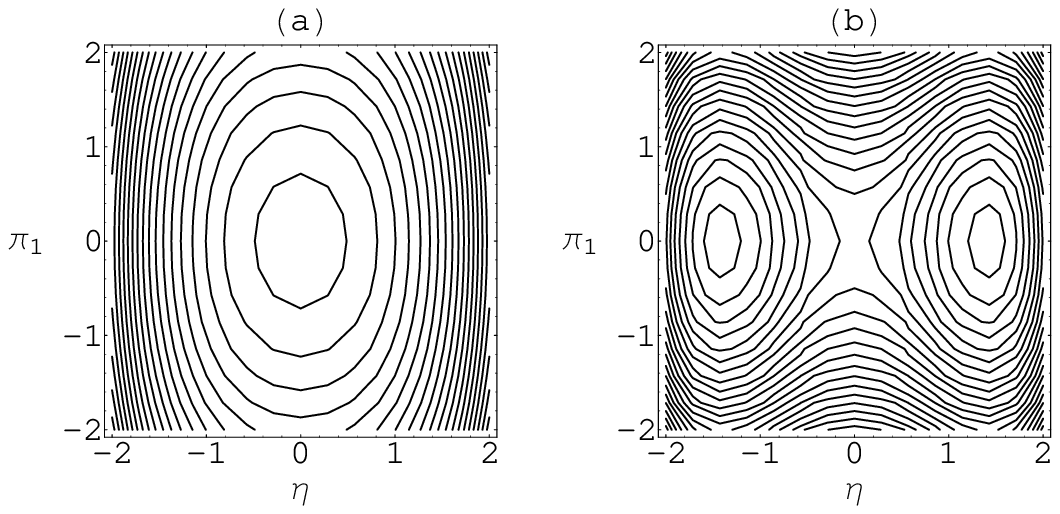}}
\caption{\label{fig:5}Pictorial representation of (a) hard spring case and (b)
soft spring case. Note that in the soft spring case we have two
instability points $\pi_1=\pm\frac{\sqrt{2}M}{\lambda}$.}
\end{figure*}

Direct integration of this equation results

\begin{eqnarray}
\label{10} \eta^2+\frac{\lambda^2}{2}{\pi_1}^4\pm2M^2{\pi_1}^2=2c,
\end{eqnarray}

\noindent where $c$ is an integration constant with dimensions of
$\mathrm{MeV}^4$. Figure \ref{fig:5} show a pictorial
representation of $(\eta,\pi_1)$ phase space. In Duffing's theory
context, this $c$ parameter can be interpreted as an energy level
to a given closed trajectory. For the case
$\eta^2+\frac{\lambda^2}{2}{\pi_1}^4+2M^2{\pi_1}^2=2c$, we get the
so-called hard spring case. The solutions are closed trajectories
centered at the origin of the coordinate system, as showed in
Figure \ref{fig:5}a. In this case, only positive values for $c$
are permitted. For
$\eta^2+\frac{\lambda^2}{2}{\pi_1}^4-2M^2{\pi_1}^2=2c$, we have
the soft spring case and, like the previous one, the closed
trajectories, which exist only near the center of the potential,
represents yet periodic solutions. For these situations, each
value of $c$ represents an energy level for a given trajectory, as
showed in Figure \ref{fig:5}b. In this case, far from the center
of the potential, the solutions are non-periodic, for an arbitrary
but fixed value of $c$ \cite{struble}.

For the hard spring case we have only one real minimum at
$(\eta=0,\pi_1=0)$. For the soft spring case the center of the
coordinate system is an instability point given by
$(\eta=\pm\sqrt{c},\pi_1=0)$ which separates two maxima, each one
localized near of the system center and are given by
$\pi_1=\pm\frac{\sqrt{2}M}{\lambda}$ \cite{struble}.

A separation of variables followed by an integration take equation
(\ref{10}), for both cases ($+2M^2$ and $-2M^2$), into the form:

\begin{eqnarray}
\label{11}
\int^{z}_{0}\frac{dz'}{\sqrt{-z'^4{\pm}z'^2+\frac{p}{2}}}=\sqrt{2}Mt,
\end{eqnarray}

\noindent where, for the sake of simplicity, we took $t_0=0$, and
defined $z=\frac{\lambda}{2M}{\pi_1}$ and
$\frac{p}{2}=\frac{\lambda^2{c}}{4M^4}$.

The above integral depends strongly on the values of $p$. So we
split it in several cases as showed below.
\subsubsection*{Exact Solutions: Case $-2M^2$}
$\mathbf{p<-1/2}$: In this case we do not have real solutions
for the integral presents in (\ref{11}).

$\mathbf{p>0}$: It is possible to write the roots of $-z'^4-z'^2+\frac{p}{2}=0$ as

\begin{eqnarray}
\nonumber r^2=-\frac{1}{2}+\frac{1}{2}\sqrt{1+2p}, \hspace{0.3cm}s^2=+\frac{1}{2}+\frac{1}{2}\sqrt{1+2p},
\end{eqnarray}

\noindent and using the table of integrals in \cite{grad} we write
the integral in (\ref{11}) as

\begin{eqnarray}
\label{a1}
\int^{z}_{0}\frac{dz'}{\sqrt{(z'^2+r^2)(s^2-z'^2)}}=\frac{1}{\sqrt{r^2+s^2}}K[\gamma,q^2],
\end{eqnarray}

\noindent where $K[\gamma,q^2]$ is the elliptic integral of first kind
given by

\begin{eqnarray}
\nonumber
\int^{\gamma'}_{0}\frac{d\gamma}{\sqrt{(1-q^2)(1-\sin^2{\gamma})}},
\end{eqnarray}

\noindent and

\begin{eqnarray}
\nonumber
\gamma=\arcsin\left(\frac{z}{s}\sqrt{\frac{r^2+s^2}{r^2+z^2}}\right),\hspace{0.3cm}q^2=\frac{s^2}{r^2+s^2},
\end{eqnarray}

\noindent and $q^2$ is the elliptic function modulus ($0\leqslant{q}\leqslant{1}$).

From (\ref{11}) and (\ref{a1}) and after some algebra one obtains

\begin{eqnarray}
\label{12}\frac{\lambda}{2M}{\pi_1}=\frac{\sqrt{2p}sn}{\sqrt{\sqrt{1+2p}(2+sn^2)-sn^2}},
\end{eqnarray}

\noindent where, for short,

\begin{eqnarray}
\nonumber
sn=sn\left[[4(1+2p)]^{\frac{1}{4}}Mt,\frac{1+\sqrt{1+2p}}{2\sqrt{1+2p}}\right],
\end{eqnarray}

\noindent is the Jacobi's elliptic sine that possess a double
period. The behavior of this solution can be fixed by an
appropriate choice of the parameters $M$ and $p$.

$\mathbf{-1/2{\leq}p{\leq}0}$: Again, using \cite{grad} we obtain two
distinct solutions which depends on the roots

\begin{eqnarray}
\nonumber r^2=+\frac{1}{2}+\frac{1}{2}\sqrt{1+2p},\hspace{0.3cm}s^2=+\frac{1}{2}-\frac{1}{2}\sqrt{1+2p},
\end{eqnarray}

Using \cite{grad} we can write

\begin{eqnarray}
\label{a2}
\int^{z}_{0}\frac{dz'}{\sqrt{-z'^4{\pm}z'^2+\frac{p}{2}}}=\frac{1}{r}K[\zeta,q^2],
\end{eqnarray}

\noindent where now

\begin{eqnarray}
\nonumber
\zeta=\arcsin\left(\frac{r}{z}\sqrt{\frac{z^2-s^2}{r^2-s^2}}\right),\hspace{0.3cm}q^2=\frac{r^2-s^2}{r^2}.
\end{eqnarray}

Again, from equations (\ref{a2}) and (\ref{11}) we get

\begin{eqnarray}
\label{13}\frac{\lambda}{2M}{\pi_1}=\frac{\sqrt{-2p}}{\sqrt{1+\sqrt{1+2p}(1-2sn^2)}},
\end{eqnarray}

Using \cite{grad} and writing the roots as done in the case above one obtains

\begin{eqnarray}
\label{a3}
\int^{z}_{0}\frac{dz'}{\sqrt{-z'^4{\pm}z'^2+\frac{p}{2}}}=\frac{1}{r}K[\varphi,q^2],
\end{eqnarray}

\noindent where

\begin{eqnarray}
\nonumber
\varphi=\arcsin\left(\sqrt{\frac{r^2-z^2}{r^2-s^2}}\right),\hspace{0.3cm}q^2=\frac{r^2-s^2}{r^2}.
\end{eqnarray}

Finally, from equations (\ref{a3}) and (\ref{11}) one obtains

\begin{eqnarray}
\label{14}\frac{\lambda}{2M}{\pi_1}=\sqrt{1+\sqrt{1+2p}(1-2sn^2)},
\end{eqnarray}

\noindent where, in both cases,

\begin{eqnarray}
\nonumber sn=sn\left[\sqrt{1+\sqrt{1+2p}}Mt,\frac{2\sqrt{1+2p}}{1+\sqrt{1+2p}}\right].
\end{eqnarray}

Again, all characteristics of these solutions are controlled only
by the free parameters $M$ and $p$.

\subsubsection*{Exact Solutions: Case $+2M^2$}

In the case $+2M^2$, using \cite{grad}, we obtain just
only one solution.

$\mathbf{p>0}$: We can write the roots of $-z'^4+z'^2+\frac{p}{2}=0$ as

\begin{eqnarray}
\nonumber r^2=-\frac{1}{2}+\frac{1}{2}\sqrt{1+2p},\hspace{0.3cm}s^2=+\frac{1}{2}+\frac{1}{2}\sqrt{1+2p},
\end{eqnarray}

\noindent and we obtain the same solution expressed by equation
(\ref{12}). Solutions obtained until now does not possess
singularities and have a time periodic evolution. Jacobian sine
behaves, roughly speaking, as a trigonometric sine.

In our numerical analysis we noted that for an appropriate fixed
$M$ (140 $\mathrm{MeV}$ or 420 $\mathrm{MeV}$), frequencies
oscillation versus $p$ of equations in (\ref{12}), (\ref{13}),
(\ref{14}) presents a logarithm form, that is, for some fixed
value of $p$, the frequency have a slow, but growth behavior. The
PR phenomenon here depends on frequency oscillation of $\pi^-$
field and it indicates that, from a determined value of $p$, the
particle creation rate itself have a slow, but continuous growth.

\subsection{Inhomogeneous Case (${|\Psi_n|}^2{\neq}0$)}

The $|\Psi_n|^2$ field will be treated here, for the sake of
simplicity, as an isotropic, homogeneous medium where interactions
between the pion condensate and the $A^{\mu}$ field can take
place. In other words, $|\Psi_n|^2=\Psi^{\dag}_n \Psi_n$ possess a
constant value in the NS medium.

Here, $|\Psi_n|^2$ describe a neutron matter in its condensate
phase. The condensation of neutrons can take place in the usual
way, the neutrons comprise Cooper pairs which obey the
Bose-Einstein statistics. In the condensed phase its
ground state assume a non-zero value, that is

\begin{eqnarray}
\nonumber {\langle{\Psi_n}\rangle}_{vac}\neq{0}.
\end{eqnarray}

In this phase it is well-known that neutrons can be described by a
standing wave function. We take the approximation
$\Psi_n=\langle{\Psi_n}\rangle$ and write

\begin{eqnarray}
\nonumber
\langle{\Psi_n}\rangle={\langle{\Psi_n}\rangle}_{vac}{e^{i\vec{k}{\cdot}\vec{r}}},
\end{eqnarray}

\noindent where $\vec{k}$ is the momentum of neutron condensate.
If we assume isotropy of the condensate neutron we can write it as

\begin{eqnarray}
\nonumber
\langle{\Psi_n}\rangle={\langle{\Psi_n}\rangle}_{vac}{\exp[ikr]}.
\end{eqnarray}

Taking into account this result and using the
relation (\ref{2}) we obtain a nonlinear second order
non-homogeneous system

\begin{widetext}
\begin{equation}
\left\{\begin{array}{r@{\quad\quad}} \label{16}
\frac{d^2}{dt^2}\pi_1\pm{2M^2}\pi_1+\frac{\lambda^2}{2}{\pi_1}^3+\frac{\lambda^2}{2}{\pi_2}^2{\pi_1}=-g{{\langle{\Psi_n}\rangle}^2}_{vac}{\sin(2kr)}\\
\frac{d^2}{dt^2}\pi_2\pm{2M^2}\pi_2+\frac{\lambda^2}{2}{\pi_2}^3+\frac{\lambda^2}{2}{\pi_1}^2{\pi_2}=-g{{\langle{\Psi_n}\rangle}^2}_{vac}{\cos(2kr),}
\end{array}\right.
\end{equation}
\end{widetext}

\noindent where $g$ is the coupling constant presents in the
interaction Lagrangian. In order to get a qualitative study we
take $h=g{\langle{\Psi_n}\rangle^2}_{vac}$ as a free parameter and
analyze the behavior of the pion condensate for some values of
$h$.

\begin{figure*}
\centering
\includegraphics[scale=0.8]{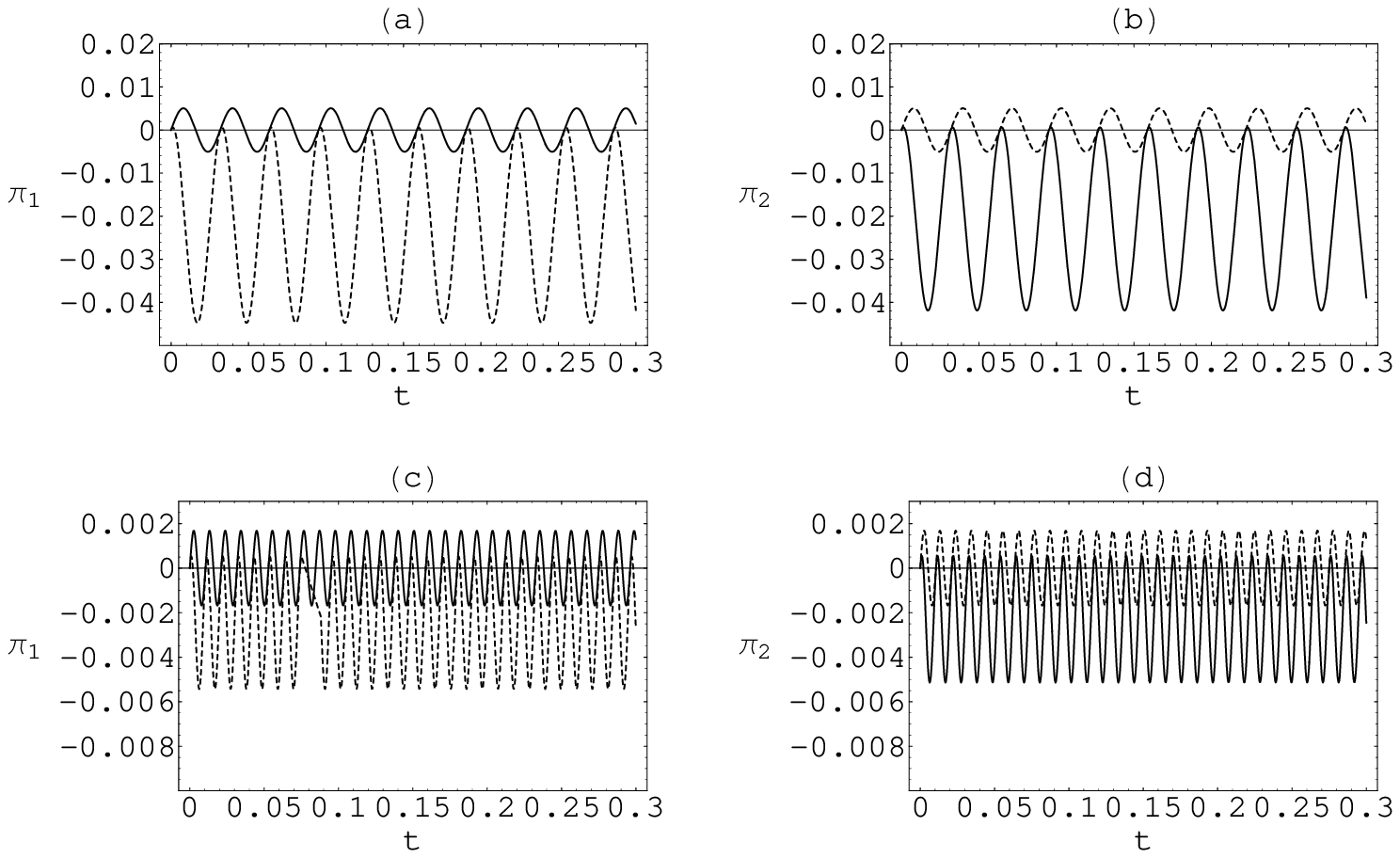}
\caption{\label{fig:6}Figures above show the behavior of $\pi_1$
and $\pi_2$ for the case $-2M^2$. For (a) and (b) $M=140$
$\mathrm{MeV}$: continuous-line ($h=1000$) and dashed-line
($h=0.001$). For (c) and (d) $M=420$ $\mathrm{MeV}$:
continuous-line ($h=1000$) and dashed-line ($h=0.001$). Note that
even under extreme variation of $h$ the regularity still persists
($h=g{\langle{\Psi_n}_{vac}\rangle^2}$).}
\end{figure*}

\begin{figure*}
\centering
\includegraphics[scale=0.8]{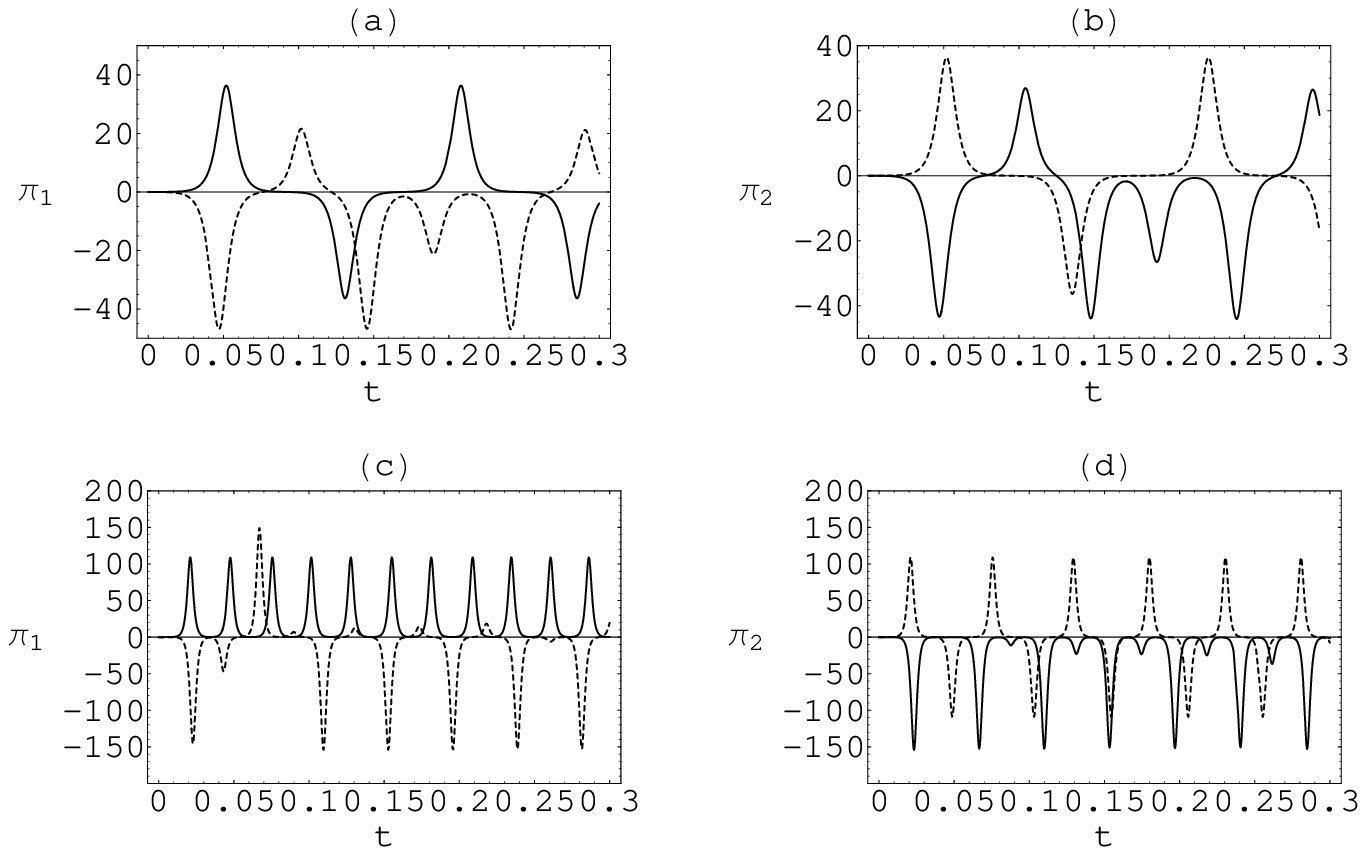}
\caption{\label{fig:7}Figures above show the behavior of $\pi_1$
and $\pi_2$ for the case $+2M^2$. For (a) and (b) $M=140$
$\mathrm{MeV}$: continuous-line ($h=1000$) and dashed-line
($h=0.001$). For (c) and (d) $M=420$ $\mathrm{MeV}$:
continuous-line ($h=1000$) and dashed-line ($h=0.001$). Note that
even under extreme variation of $h$ the regularity still persists
($h=g{\langle{\Psi_n}_{vac}\rangle^2}$).}
\end{figure*}

Virtually, the above system does not have an exact solution. So,
we can make only a numerical study of its temporal evolution.
Figure \ref{fig:6} shows the temporal behavior of $\pi_1$ (a) and
$\pi_2$ (b) for ($-2M^2$). Note that regularity persists even for
a wide range of $h$; Figure \ref{fig:7} shows the time behavior of
$\pi_1$ (a) and $\pi_2$ (b) for ($+2M^2$). Again, the regularity
persists. Some irregularities in the figures are due to numerical
approximation.

In this paper we will only analyze below the PR phenomenon through
exact solutions of the pion condensate EoM. The fact that
inhomogeneous system (\ref{16}) possess periodic solutions
suggests that the PR phenomenon can persists even in this
situation.

\section{Photoproduction by the $A^{\mu}$ field oscillations}

In this section we study solutions obtained for the time dependent
electromagnetic field coupled to the pion condensate field. From the
Lagrangian density (\ref{1}) one obtains the EoM of the electromagnetic
field which reads:

\begin{widetext}
\begin{eqnarray}
\label{17} \Box{A^{\mu}}-\partial^{\mu}\partial_{\nu}A^{\nu}+ie(\pi^+\partial^{\mu}\pi^--\pi^-\partial^{\mu}\pi^+)+2e^2(\pi^-\pi^+)A^{\mu}-e\gamma^{\mu}\frac{1}{2}\bar{\Psi}(1+\tau_3)\Psi=0.
\end{eqnarray}
\end{widetext}

Note that the interaction term $2e^2(\pi^-\pi^+)A^{\mu}$ is the
most interesting in the above equation to our purposes. As we will
see in a moment below, it is responsible for the PR phenomenon. We
need to make some approximations in order to get an EoM which can
be handled more easily. We start by neglecting nucleon
contributions, i.e., we firstly study the Homogeneous Case
(${|\Psi_n|}^2=0$).  This can be made to first order approach
\cite{hs1,hs2}. In the next subsection we study this case.

\subsection{Photoproduction in the Homogeneous Case (${|\Psi_n|}^2=0$)}

For ${|\Psi_n|}^2=0$ the equation (\ref{17}) reads

\begin{eqnarray}
\nonumber\Box{A^{\mu}}-\partial^{\mu}\partial_{\nu}A^{\nu}+ie(\pi^+\partial^{\mu}\pi^--\pi^-\partial^{\mu}\pi^+)+\\
\nonumber +2e^2(\pi^-\pi^+)A^{\mu}=0.
\end{eqnarray}

Now, from the symmetry of pion system in (8) we can assume, 
for the sake of simplicity, 
the ansatz $\pi_2=\pi_1$ the term
$ie(\pi^+\partial^{\mu}\pi^--\pi^-\partial^{\mu}\pi^+)$ vanishes.
This ansatz was suggested by the analysis done above of the
solutions for the pion condensate solutions obtained from the
homogeneous system for the case ${|\Psi_n|}^2=0$. Thus, the above
equation reduces to

\begin{eqnarray}
\label{18}\Box{A^{\mu}}-\partial^{\mu}\partial_{\nu}A^{\nu}+2e^2(\pi^-\pi^+)A^{\mu}=0.
\end{eqnarray}

Introducing the gauge field transformation

\begin{eqnarray}
\label{19} A^\mu{\rightarrow}{A'}^\mu=A^\mu+\partial^\mu\zeta(\vec{r},t),
\end{eqnarray}

\noindent where $\zeta(\vec{r},t)$ is an arbitrary function of
$\vec{r}$ and $t$ and using the Lorentz gauge condition

\begin{eqnarray}
\nonumber\partial_{\mu}{A'}^\mu=\partial_\mu{A^\mu}+\Box\zeta(\vec{r},t)=0,
\end{eqnarray}

\noindent since $-\partial^{\mu}\partial_{\nu}A^{\nu}$ vanishes. An isotropic gauge field ${A'}^{\mu}$ can be written as

\begin{eqnarray}
\label{b1} {A'}^\mu(x)=\chi(x)e^\mu],
\end{eqnarray}

\noindent where $\chi(x)$ can be expanded in Fourier modes and
$e^\mu$ is a polarization vector.

Substituting (\ref{b1}) in (\ref{18}) and taking into account the
Lorentz gauge condition one obtains

\begin{eqnarray}
\label{20}e^\mu[\Box\chi(x)+2e^2(\pi^-\pi^+)\chi(x)]=0.
\end{eqnarray}

Now, in a crude, but nevertheless unrealistic approximation, we
can take also the electromagnetic field as homogeneous inside the
NS. So (\ref{20}) reduces to an ordinary differential equation in
the time variable. Taking the Fourier transform to $k$ space
momentum  we finally get

\begin{eqnarray}
\label{21}\frac{d^2}{dt^2}\chi_k(t)+[k^2+2e^2(\pi^-\pi^+)]\chi_k(t)=0.
\end{eqnarray}

This is a homogeneous differential equation for an oscillator with
frequency (or mass) varying in time. We define the frequency
oscillation of the $\chi_k{(t)}$ field as

\begin{eqnarray}
\nonumber \omega_k=k^2+2e^2(\pi^-\pi^+).
\end{eqnarray}

For the Homogeneous Case (${|\Psi_n|}^2=0$) and solutions of
${\pi^-}$ and ${\pi^+}$ fields, previously obtained, we may write
(\ref{21}) as a Lam\'{e}'s type equation.

\begin{figure*}
\centering
\includegraphics[scale=1.1]{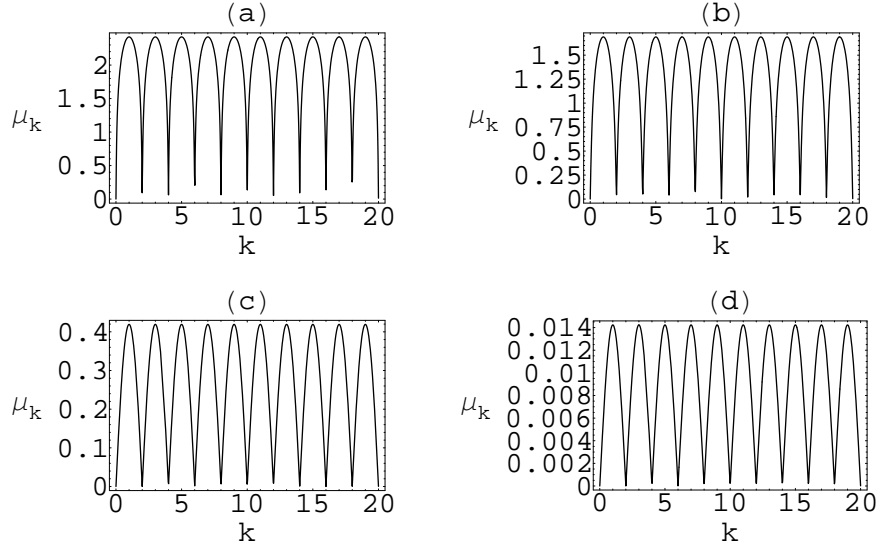}
\caption{\label{fig:8}Floquet exponent $\mu_k$ as a function of $k$ and fixed
values of the determinant $\Delta(0)$: (a) $\Delta(0)=-1000$, (b)
$\Delta(0)=-100$, (c) $\Delta(0)=-1$ and (d) $\Delta(0)=-0.001$. If
$\Delta(0)<0$, $\mu_k$ is real; if $\Delta(0)>0$, $\mu_k$ is imaginary.}
\end{figure*}

A general Lam\'e's Equation have the form

\begin{eqnarray}
\nonumber \frac{d^2}{dt^2}\psi(t)+[A+BQ(t)]\psi(t)=0,
\end{eqnarray}

\noindent where $A$ and $B$ are real numbers and $Q(t)$ is an
elliptic function. Phase space of the parameters $A$ and $B$ is
not easily defined, and this is exactly the case of equation
(\ref{21}). Nevertheless it still possess an important property,
namely, the existence of instability (exponential growth)

\begin{eqnarray}
\label{22} \chi_k(t){\sim}\exp[{\mu_k}t],
\end{eqnarray}

\noindent where ${\mu_k}\in(-\infty,\infty)$ is the so-called
Floquet exponent (or characteristic exponent) labelled by the $k$
momentum. To find exact solutions for the Lam\'{e}'s Equation is a
very difficult task. For example, if the coefficient of
$(\pi^-\pi^+)$ term in (\ref{21}) may be written as $n(n+1)q^2$,
where $n$ is an integer and $0{\leqslant}q{\leqslant}1$ is the
elliptic function modulus, then an exact solution can be founded.
On the other hand for a qualitative study we can use $\mu_k$ taken
from the formal solution of Hill's Equation \cite{arscott}, which
can be written as

\begin{eqnarray}
\label{23}\mu_k=\frac{1}{\pi}\cosh^{-1}[1-\Delta(0)\sin^2(\frac{k{\pi}}{2})],
\end{eqnarray}

\noindent where $\Delta(0)$ is an infinite determinant. Figure
\ref{fig:8} shows $\mu_k>0$ for arbitrary values of $k>0$ momentum
with several fixed $\Delta(0)$ values.

The Floquet's theorem \cite{win} gives

\begin{eqnarray}
\nonumber \chi_k(t)=\exp[{{\mu_k}t}]p(t),
\end{eqnarray}

\noindent where $p(t)=p(t+T)$ is a real limited periodic function
with period $T$. When ${\mu_k}>0$, instability corresponds to an
exponential growth of occupation number 

\begin{eqnarray}
\label{xx}n_k(t){\sim}\exp[2{\mu_k}t],
\end{eqnarray}

\noindent that may be interpreted as number density of particles
produced by PR mechanism \cite{dk,kls}.

In the following subsections will analyze the equation (\ref{21})
by means solutions (\ref{12}), (\ref{13}) and (\ref{14})
for the $\pi^-$ condensate.

\subsubsection{Homogeneous (${|\Psi_n|}^2=0$): Case $-2M^2$}

In order to obtain the Lam\'{e}'s Equations, the solutions
obtained for the pion condensate EoM were substituted in equation
(\ref{21}). All solutions that we found for the nonlinear field
(source) are periodic and substituting them in (\ref{21}) we got a
Hill's Equation. For the sake of simplicity, below we show some of
them.


Using equation (\ref{12}) and after some simplifications we write
(\ref{21}) as

\begin{widetext}
\begin{eqnarray}
\label{x0}\frac{d^2}{dt^2}{\chi_k}(t)+\left[k^2+\frac{16M^2e^2p}{\lambda^2[\sqrt{1+2p}(2+sn^2)-sn^2]}sn^2\right]{\chi_k}(t)=0,
\end{eqnarray}
\end{widetext}

\noindent where

\begin{eqnarray}
\nonumber\omega_k=k^2+\frac{16M^2e^2psn^2}{\lambda^2[\sqrt{1+2p}(2+sn^2)-sn^2]},
\end{eqnarray}

The above equation may be classified as a Lam\'{e}'s type having a
Jacobian periodicity, and, \textit{a priori}, may be write in a
more handle way, i.e., as just one Jacobian sine or a sum of them.
For our purposes, it is sufficient in the Hill's theory that the
potential (\ref{12}) has a periodic behavior.


Using solution (\ref{13}), equation (\ref{21}) can be written as

\begin{widetext}
\begin{eqnarray}
\label{x1} \frac{d^2}{dt^2}{\chi_k}(t)+\left[k^2+\frac{16M^2e^2p}{\lambda^2[1+\sqrt{1+2p}(1-2sn^2)]}\right]{\chi_k}(t)=0,
\end{eqnarray}
\end{widetext}

\noindent where clearly

\begin{eqnarray}
\nonumber\omega_k=k^2+\frac{16M^2e^2p}{\lambda^2[1+\sqrt{1+2p}(1-2sn^2)]}.
\end{eqnarray}

We got again an equation whose potential is written in terms of
elliptic functions and it is of Lam\'e's type.


Now, using (\ref{14}), equation (\ref{21}) can be written as

\begin{widetext}
\begin{eqnarray}
\label{x2} \frac{d^2}{dt^2}{\chi_k}(t)+
\left[k^2+\frac{8M^2e^2(1+\sqrt{1+2p})}{\lambda^2}-
\frac{16M^2e^2\sqrt{1+2p}}{\lambda^2}sn^2\right]{\chi_k}(t)=0,
\end{eqnarray}
\end{widetext}

with

\begin{eqnarray}
\nonumber\omega_k=k^2+\frac{8M^2e^2(1+\sqrt{1+2p})}{\lambda^2}-\frac{16M^2e^2\sqrt{1+2p}}{\lambda^2}sn^2.
\end{eqnarray}

This is, clearly, a Lam\'e's Equation which is easier to handle than
the another ones above.

\subsubsection{Homogeneous (${|\Psi_n|}^2=0$): Case $+2M^2$}

In this case we get just one equation. Using (\ref{12}) equation
(\ref{21}) may be written as

\begin{widetext}
\begin{eqnarray}
\label{x3} \frac{d^2}{dt^2}{\chi_k}(t)+\left[k^2+\frac{16M^2e^2p}{\lambda^2[\sqrt{1+2p}(2+sn^2)-sn^2]}sn^2\right]{\chi_k}(t)=0,
\end{eqnarray}
\end{widetext}

\noindent where

\begin{eqnarray}
\nonumber\omega_k=k^2+\frac{16M^2e^2p}{\lambda^2[\sqrt{1+2p}(2+sn^2)-sn^2]}sn^2,
\end{eqnarray}

\noindent represents its mass oscillation coefficient.

All Lam\'{e}'s Equations obtained above have only two free
parameters, $M$ and $p$. These two free parameters control
frequency and amplitude of pion oscillations and consequently  by
equation (\ref{xx}) also the rate of particle production of the
photon field.

Energy density for each Lam\'{e}'s Equation can be calculated by
\cite{gbk}

\begin{eqnarray}
\label{x6}\rho_{\gamma}=\frac{1}{2\pi^2}{\int_0}^{k_c}dk\omega_k{n_k(t)}k^2.
\end{eqnarray}

We are not taking into account here the back reaction of the particles
produced.

In order to obtain a more realistic physical description is
necessary introduce by hand a term, in (\ref{1}) or, yet
in (\ref{21}), that represents the back reaction of
particles produced. Since this work aims just to
propose a new process of photoproduction in NS, it is sufficient to
prove that the PR phenomenon could exist in the NS medium, as it is shown above.

\subsection{Photoproduction in the Inhomogeneous Case (${|\Psi_n|}^2{\neq}0$)}

Using the gauge transformation (\ref{19}) with the Lorentz
condition, equation (\ref{17}) can be written as

\begin{widetext}
\begin{eqnarray}
\label{33} \Box{A^{\mu}}+ie(\pi^+\partial^{\mu}\pi^--\pi^-\partial^{\mu}\pi^+)+2e^2(\pi^-\pi^+)A^{\mu}-e\gamma^{\mu}\frac{1}{2}\bar{\Psi}(1+\tau_3)\Psi=0.
\end{eqnarray}
\end{widetext}

Now we take as zeroth approximation the nucleons contribution as a constant in space and time variables. This is, of course, a crude approximation relied on isotropy, homogeneity  and independence on time for the nucleon  matter inside the star core. Nevertheless the figures 
\ref{fig:6} and \ref{fig:7} shows that periodicity of the pion fields is robust against extreme variations of the nucleons contributions (in zeroth order). 
That is, the nucleons have only a weak influence on the pion evolution, whereas the strong influence on amplitude and frequency comes from the value and signal of pion mass term $M$. 
In adition we believe that so restrict conditions of constance of nucleons contribution can be relaxed and the periodicity of pion fields will not strongly affected. This is enough in order to PR mechanism works well. Therefore, for our qualitative approach we can neglect the nucleon contribution. Taking the Fourier Transform, for the approximation described above,we obtain the same Lam\'e Equation (24).

%
%
%

All numerical solutions suggest that equation (24) possess
resonant bands, that is, the PR phenomenon may occurs even in the
presence of nucleons. 
Clearly, since $n_k$ has exponential behavior, so $\rho_k$ in
(\ref{x6}).

\section{Parametric Resonance and Neutrons Stars Flares}

We had showed that Parametric Resonance phenomenon, under certain
conditions, may occur inside of a NS modifying briefly its thermal
evolution. It is worth to mention that our approach can be
implemented, not only for the Harrington and Shepard's model, but
in fact for any NS model which presents superconductor phase with
periodic motions (EoM) and a suitable coupling to Electromagnetic
Field to furnish a Hill's Equation.
PR mechanism can be effective enough to produce a huge
number of photons, out of equilibrium with the surrounding matter, which must
diffuse out from the NS. In this case, clearly, a subsequent
reheating of the NS interior must occur, much the same as in the
Inflationary Theory, but in a far smaller energy scale. 
From the mathemartical point of view we have two cases to 
consider, namely, for real Floquet exponent and when it is an imaginary number. 
Both cases lead to an enhancing of the NS brightness, but the possible observable effects are expected to be very different as detailed below. 

Taking for granted the current theories of NS interiors as nearly correct,
in our opinion, this kind of PR phenomenon can, in principle, be detected by
observation of anomalous enhancing of the NS brightness. The
intensity, as well the lapse of time of this higher brightness
depends on the details of diffusion of the produced photons to reach the surface of
the star. This is a difficult problem since, by its turn, it depends on the
model of the NS matter structure.
On the other hand, recent works  seem to indicate a nearly black body surface
radiation from Neutron Stars  whose details also depends on the model
of their atmospheres \cite{blackbody,spectrum1,spectrum2,spectrum3}. If this is correct, in the case that PR mechanism is active, it leads to some interesting results on the radiation released by a NS. We consider first the case of real Floquet exponent.
On a blackbody spectrum background, PR  manifests its presence, by the theory outlined above, in a weak regime, as a
perturbation on NS spectrum with superimposed bumps on the equally
spaced frequencies due to the periodic pattern of the Floquet
exponent as a function of the momentum $k$, as is clearly seen in
equation (\ref{23}) and in Fig. \ref{fig:8}. This periodicity implies that
high energy photons are also produced and their spectral lines should 
dominate over the blackbody ones, which lead us to expect that PR effects could be better measured on the tails of a blackbody spectrum. 
Furthermore, although in a qualitative ground, in the case of a very
strong regime (with a large real Floquet exponent) of photon production,  
the external star layers  can not
resist against the enormous pressure due to the reheating  of
matter in the star core by high energy photons. In this case
the star ejects its upper layers with a huge burst of energy releasing. 
In this way our model suggest a feasible and simple 
explanation for the recent observations of neutron stars giant flares \cite{explosion1,explosion2,explosion3}. Shortly, PR mechanism with a real and large  Floquet exponent is better to explain non-periodic, but giant flares from NS.
On the other hand, if the Floquet exponent is imaginary, Eq. (27) shows that the number density of produced particles is a periodic function in time and so we can expect a periodic pertubation on the brightness of the NS while the PR regime is active. This is quite interesting since giant periodic flares 
from repeaters have been observed in the past few years. See, for example, \cite{explosion3}.
However, for this case, we can have only an incomplete view since for giant flares  the unkown coefficient in Eq.(27) must be very large, since the time factor is a trigonometrical function and so a bounded function. So, a further study is needed to solve this quest.

\section{Conclusion}
Firstly, it must be stressed that Parametric Resonance in NS can be
a rare phenomenon, due to the special conditions of the
superconductor matter and its coupling with the electromagnetic field.
Nevertheless, in the weak PR regime, as mentioned above, 
it is worth to notice that, in principle, it is possible to measure  an uprising regular (equally spaced frequencies) brightness variation on the nearly blackbody spectra, perhaps better measured on their tails, of those neutron stars close enough to us. In adition, in the strong PR regime, the model offers also a simple explanation for some reported rare giant flares coming from these compact objects. 
On the other hand  the signal can also presents a time periodic pattern. In this case, as in Inflationary Theory, it is not expected an explosive behaviour of the star, but only periodic flares with low intensity. 

Finally, more importantly perhaps, we are inclined to think that Parametric Resonance is a phenomenon active not only in Inflationary Cosmology but in several much smaller coupled physical systems with extreme conditions of pressure and temperature in our Universe.

\section*{Acknowledgments}

The authors are in debt to Prof. Waldyr A. Rodrigues Jr. and Prof.
Robert H. Brandenberger for useful comments. S.D.C. was supported
by Conselho de Desenvolvimento Cient\'ifico e Tecnol\'ogico (CNPq), Brazil. 
A.M.J was supported  by Funda\c{c}\~ao de Amaparo \`a Pesquisa do Estado de S\~ao Paulo (FAPESP) and Coordenadoria de Aperfei\c{c}oamento de Pessoal de Ensino Superior (CAPES), Brazil.


\end{document}